\documentclass[transmag]{ieeeoj_minimalheading}

\usepackage[utf8]{inputenc}
\usepackage[T1]{fontenc}
\usepackage[english]{babel}
\usepackage{microtype}
\usepackage{blindtext}

\usepackage[style=authoryear,sorting=none,natbib=true,backend=biber]{biblatex}
\usepackage[pdftex]{graphicx}
\graphicspath{{../../fig/}{../../build/}}

\usepackage{pgfplots}
\pgfplotsset{compat=1.16}

\usepackage[acronym]{glossaries}
\newglossaryentry{lui}
{
    name=\mbox{Lu.i},
    description={Educational neuron PCB implementing LIF dynamics.}
}
\newglossaryentry{cuba}
{
    name={current-based},
    description={type of synapse model}
}

\newacronym{pcb}{PCB}{printed circuit board}
\newacronym{lif}{LIF}{leaky-integrate-and-fire}
\newacronym{psp}{PSP}{postsynaptic potential}
\newacronym{snn}{SNN}{spiking neural network}

\usepackage{nicefrac}
\usepackage{placeins}  %
\usepackage{csquotes}
\usepackage{scalerel}

\usepackage{ragged2e}
\usepackage{flushend}  %

\usepackage{siunitx}
\newcommand*\subtxt[1]{_{\textnormal{#1}}}
\DeclareRobustCommand\_{\ifmmode\expandafter\subtxt\else\textunderscore\fi}

\PassOptionsToPackage{hyphens}{url}
\usepackage[
    hidelinks,
    pdftitle={Lu.i -- A low-cost electronic neuron for education and outreach},
    pdfauthor={Yannik Stradmann, Julian Göltz, Mihai A. Petrovici, Johannes Schemmel, Sebastian Billaudelle},
]{hyperref}
\usepackage[capitalise]{cleveref}

\title{%
	\vspace*{-3mm}\par %
	Lu.i -- A low-cost electronic neuron for education and outreach
}

\author{%
	Yannik Stradmann\textsuperscript{*,\,1},
	Julian Göltz\textsuperscript{*,\,1,\,2},\\
    Mihai A. Petrovici\textsuperscript{\,2},
    Johannes Schemmel\textsuperscript{\,1},
	Sebastian Billaudelle\textsuperscript{\,1}

	\vspace{1mm}
	{\normalfont\small
	\textsuperscript{*}\,These authors contributed equally, listed in reverse alphabetical order.

	\textsuperscript{1}\,Kirchhoff-Institute for Physics, Heidelberg University.\\[-1mm]
	\textsuperscript{2}\,Department of Physiology, University of Bern.
	}
}

\begin{document}

\begin{abstract}
With an increasing presence of science throughout all parts of society, there is a rising expectation for researchers to effectively communicate their work and, equally, for teachers to discuss contemporary findings in their classrooms.
While the community can resort to an established set of teaching aids for the fundamental concepts of most natural sciences, there is a need for similarly illustrative experiments and demonstrators in neuroscience.
We therefore introduce \gls{lui}: a parametrizable electronic implementation of the \acrlong{lif} neuron model in an engaging form factor.
These palm-sized neurons can be used to visualize and experience the dynamics of individual cells and small \acrlongpl{snn}.
When stimulated with real or simulated sensory input, \gls{lui} demonstrates brain-inspired information processing in the hands of a student.
As such, it is actively used at workshops, in classrooms, and for science communication.
As a versatile tool for teaching and outreach, \gls{lui} nurtures the comprehension of neuroscience research and neuromorphic engineering among future generations of scientists and in the general public.

\end{abstract}


\maketitle
\section{Introduction}\label{sec:introduction}

Expanding our understanding of the mammalian brain is among the central frontiers of modern science and yet implies some of the longest standing questions humanity has posed to itself.
Their fundamental nature induces an intrinsic curiosity about the progress of neuroscience, artificial intelligence, and brain-inspired technology.
In contrast to this demand, the repertoire of demonstrators to communicate principles and recent achievements in brain research is limited~\citep{gage2019case}.
In comparison, other fields can build on many centuries of experience for conveying their essential concepts.

In our current understanding, the fundamental principles of information processing in nervous systems lie in neuronal dynamics and synaptic interactions.
A strong intuition for these mechanisms is, therefore, the foundation for understanding and investigating more complex processes and emerging phenomena.
In the following, we thus present \gls{lui} -- an analog electronic implementation of the \gls{lif} neuron model targeted for educational use as well as scientific outreach.
\Gls{lui} features current-based synaptic inputs that enable the formation of simple \glspl{snn} and offers control over many parameters, including the time constants and the synaptic weights.
The \gls{pcb} visualizes the time-continuous dynamics of the emulated membrane potential and allows interfacing with digital and analog periphery for advanced experiments.
It has been optimized for low-cost production, long battery life, and intuitive operation.

\begin{figure}
	\includegraphics[width=\columnwidth]{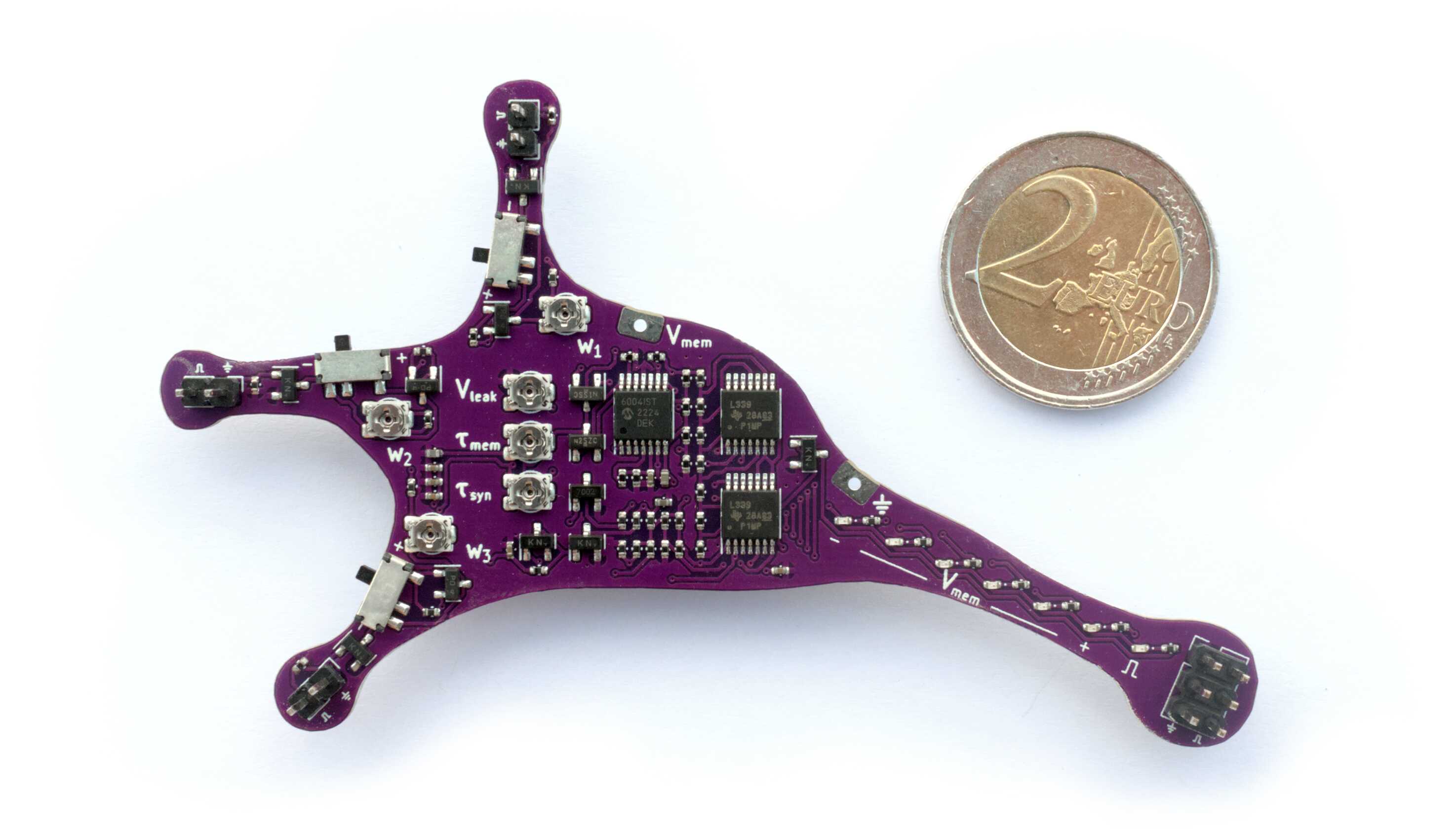}
	\caption{
		\boldmath\justifying
		A single \gls{lui} neuron \gls{pcb}, with a \mbox{2-Euro} coin for scale.
		To relay information from one neuron to the other, excitatory and inhibitory synapses can be formed by wiring the axonal output (right) to one of the three dendritic terminals (left).
		\label{fig:photograph-coin}
	}
\end{figure}

\begin{figure*}[t]
	\begin{center}
        \includegraphics[width=\textwidth]{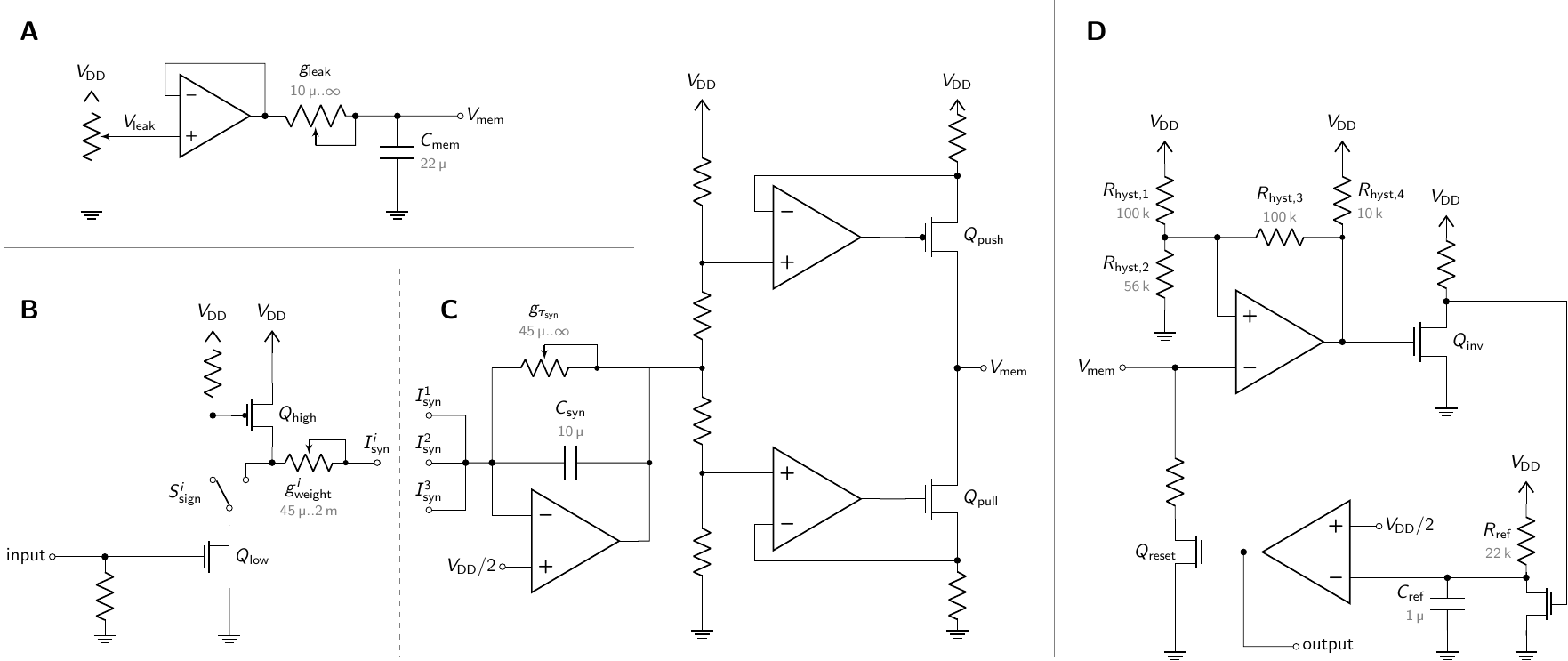}
	\end{center}
	\caption{%
        \boldmath\justifying
		Schematic of the \gls{lif} emulation circuit implemented in \gls{lui}.
		\textbf{(A)}
        Membrane capacitance and leakage resistance.
		\textbf{(B)}
        Current-based synaptic input circuits.
		This circuit is instantiated three times, once per synapse.
		\textbf{(C)}
        Synaptic integrator and voltage-to-current conversion circuit.
		\textbf{(D)}
        Threshold, reset, and refractory circuit.
		The spike output pulse is derived from the neuron's reset signal and of equivalent duration.
		\label{fig:schematic}
	}
\end{figure*}

\begin{figure*}[t]
	\includegraphics[width=\textwidth]{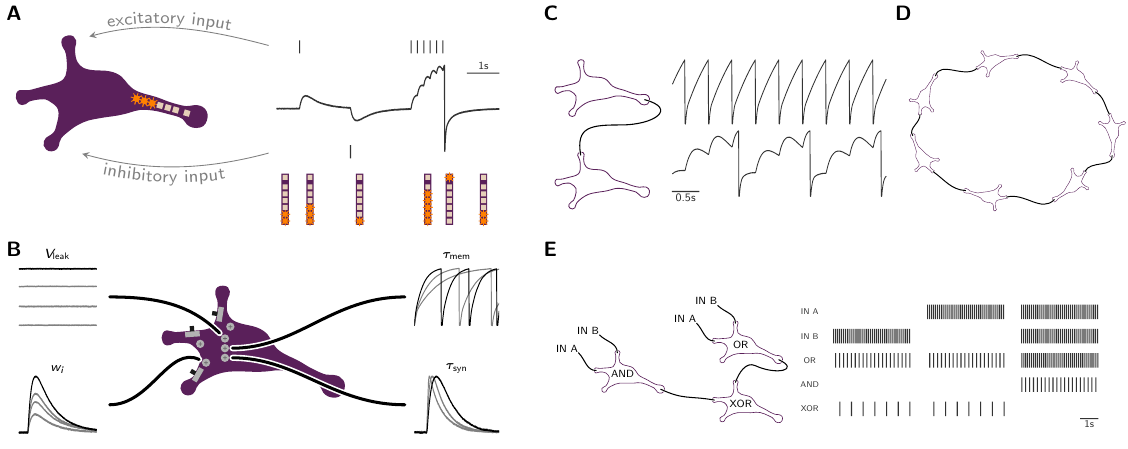}
	\caption{%
        \boldmath\justifying
		\textbf{(A)}
		A single \gls{lui} neuron receiving multiple excitatory and a single inhibitory event.
		The depicted trace shows an analog recording of $V_\text{mem}$ on the board.
		For applications without an oscilloscope at hand, each \gls{lui} neuron features a bar of LEDs to display the current membrane potential as well as axonal spikes (top LED, flashing).
		\textbf{(B)}
		Tuneable neuron parameters on \gls{lui}.
		Each model parameter is represented by a small potentiometer (cf. \cref{fig:photograph-coin}), all three synaptic weights are individually configurable in sign and strength.
		\textbf{(C)}
		Analog recording of the membrane potential $V_\text{mem}$ of two \gls{lui} neurons.
		The top trace shows the dynamics of a circuit that is configured with $V_\text{leak} > \vartheta$ and emits spikes at regular intervals.
		This neuron projects onto a second one (bottom trace), which is excited by these events, integrates the postsynaptic current and -- eventually -- also spikes.
		\textbf{(D)}
		Wiring diagram of a closed, circular delay chain built from seven \gls{lui} neurons.
		\textbf{(E)}
		Spike recording of three \gls{lui} boards, configured to represent rate-based \texttt{AND}, \texttt{OR} and -- combined -- \texttt{XOR} gates.
		The inputs \texttt{A} and \texttt{B} are presented by an external microcontroller.
		\label{fig:all_the_luis}\label{fig:psp_stacking}
	}
\end{figure*}

\section{Neuron and synapse dynamics}\label{sec:circuit}

\Gls{lui} implements the \gls{lif} neuron model, arguably the simplest abstraction that still captures the most fundamental properties of neuronal information processing: time-continuous computation, spatio-temporal integration, and event-based communication.
This model was originally put forward by Louis Lapicque in \citeyear{lapicque1907recherches}, after whom the \gls{pcb} was fittingly named.
The \gls{lif} model describes the dynamics of a neuron's membrane potential $V_\text{mem}(t)$, which are governed by the differential equation
\begin{align}
	C_\text{mem} \frac{\mathrm{d}V_\text{mem}(t)}{\mathrm{d}t} &= - g_\text{leak} \left[V_\text{mem}(t) - V_\text{leak} \right] + I_\text{syn}(t) \,,
	\label{eq:lif-dynamics}
\end{align}
where $C_\text{mem}$ denotes the membrane capacitance, $g_\text{leak}$ the leak conductance, and $V_\text{leak}$ its resting potential.
$I_\text{syn}(t)$ subsumes the time-dependent synaptic currents stimulating the neuron.
This differential equation describes a membrane potential which continuously decays to the resting state.
It is, however, augmented by a reset condition to mimick the hyperpolarization following the action potentials observed in biological neurons:
Whenever the membrane potential crosses the threshold $\vartheta$, the neuron emits a spike.
This efferent signal is accompanied by a reset of the membrane potential, where the latter is simply clamped to $V_\text{reset}$ for the refractory period.

\Gls{lui} further implements current-based synapses with postsynaptic currents following exponential kernels with time constant $\tau_\text{syn}$.
This additional temporal filter mimics the kinetics of synaptic ion channels:
Each presynaptic spike $j$, arriving at time $t_\text{pre}^j$ at synapse $i$, triggers an exponentially decaying current
\begin{align}
	I_\text{syn}^j(t) &= w_i \cdot \exp{\left(-\frac{t-t_\text{pre}^j}{\tau_\text{syn}}\right)} \,,
	\label{eq:synaptic-current}
\end{align}
where $w_i$ denotes the weight of the respective synapse $i$.
The total synaptic current then results as a sum over all of these individual contributions.

\section{Electronic implementation}\label{sec:circuit-implementation}
\Gls{lui} realizes the \gls{lif} dynamics through a set of analog electronic circuits (\cref{fig:schematic}) and thus forms a physical model thereof.
\Cref{eq:lif-dynamics} is rendered by the combination of capacitor $C_\text{mem}$ and potentiometer $g_\text{leak}$, which form an RC integrator with adjustable time constant $\tau_\text{mem}$.
Without external stimuli, $V_\text{mem}$ decays towards the resting potential $V_\text{leak}$, which we generate by the combination of an adjustable voltage divider with a subsequent unity gain buffer.
The spike mechanism is implemented by continuously comparing the membrane potential to the threshold (\cref{fig:schematic}D).
Once the membrane reaches $\vartheta=\nicefrac{V_\text{DD}}{2}$, the threshold comparator trips, indicating a spike and causing a membrane reset.
To avoid instabilities, it is fitted with a hysteresis circuit that temporarily reduces the comparator's reference potential to $\nicefrac{V_\text{DD}}{4}$ during the onset of a spike.
At that point, the capacitor $C_\text{ref}$ is discharged and the connected comparator trips, thus shorting the membrane to $V_\text{reset}=0$ via the transistor $Q_\text{reset}$ to implement the refractory period.
$R_\text{ref}$ and $C_\text{ref}$ determine the fixed refractory time of approximately \SI{12}{\milli\second}, which starts once $V_\text{mem}$ is discharged below $\nicefrac{V_\text{DD}}{4}$, where the threshold comparator releases.
The control signal for $Q_\text{reset}$ is re-used as the neuron's axonal output, with a pulse width equivalent to the refractory time.

\Gls{lui} features three synapses implementing the current-based model with an exponential kernel as introduced by \cref{eq:synaptic-current}.
Each of them possesses a tunable weight and can be switched between excitation and inhibition.
The synapses share a common synaptic time constant $\tau_\text{syn}$, which is adjustable over a broad range.
For an area- and cost-effective implementation, we minimize the amount of integrated components per synaptic connection:
Events from presynaptic neurons control the gate of the n-channel MOSFET $Q_\text{low}$ (\cref{fig:schematic}B).
Depending on the selected polarity $S_\text{sign}^{i}$, this transistor either directly discharges the shared synaptic integrator or indirectly charges it via the p-channel MOSFET $Q_\text{high}$.
For each event, this synaptic trace is in- or decremented by a fixed amount of charge proportional to the respective weight $g_\text{weight}^i$ which can be configured through a potentiometer.
The time constant $\tau_\text{syn} = C_\text{syn}/g_{\tau_\text{syn}}$ of the integrator can be similarly tuned.
Especially in light of the additional filter introduced by the membrane, this closely approximates the instantaneous response of the original model.
The synaptic current $I_\text{syn}$ is derived from the integrator state through a V-I conversion stage.
As such, it consists of two voltage-controlled current sources -- each built from a resistor, a MOSFET, and an operational amplifier.
$Q_\text{push}$ and $Q_\text{pull}$ operate in a push-pull configuration and generate two antagonistic currents.
Their difference is proportional to the deflection of the integrator and corresponds to the total postsynaptic current $I_\text{syn}$ that stimulates the membrane.

\Gls{lui} displays its state through a set of LEDs.
Six of them form a bar that visualizes the membrane potential, and a seventh LED indicates efferent spikes with a flash.
This interface is sketched in \cref{fig:all_the_luis}A for various states of the neuron.
The voltmeter is implemented through a set of comparators and a resistor ladder to generate the respective reference potentials.
While these circuits take up significant area on the \gls{pcb}, they have been omitted from the schematic for clarity.
This intuitive on-board interface enables standalone operation and the visualization of network activity and signal propagation therein.
Experimentation with external equipment is, however, encouraged and allows more detailed insights into the neuron dynamics.
For that purpose, the emulated membrane is accessible through a pad at the board edge for interfacing with, e.g., current sources and oscilloscopes.

The \gls{pcb} is powered from a single CR2032 coin cell, which we chose for its small form factor, wide availability, and comparably high capacity at low cost.
All voltage references of the circuit are derived relative to this supply.
The temporal dynamics are thus, on first order, invariant to the battery voltage.
This ensures mostly stable operation across the entire lifetime of the cell, which results in approximately \SI{24}{\hour} of continuous use.
\Gls{lui} can be powered down completely through a switch on its back side.

While aiming for an intuitive and appealing form factor, the \gls{pcb} has been strongly optimized for low-cost fabrication.
This is reflected in the selection of components as well as the layout, which only relies on a simple two-layer \gls{pcb}.
As a result, we achieved a unit price of around \SI{3}[US\$]{} already for batches below \num{1000} \gls{lui} neurons.
As the backside only contains the battery holder and an optional power switch, fabrication costs can be further reduced by restricting automated assembly to the top layer.

\begin{figure*}[t]
	\includegraphics[width=\textwidth]{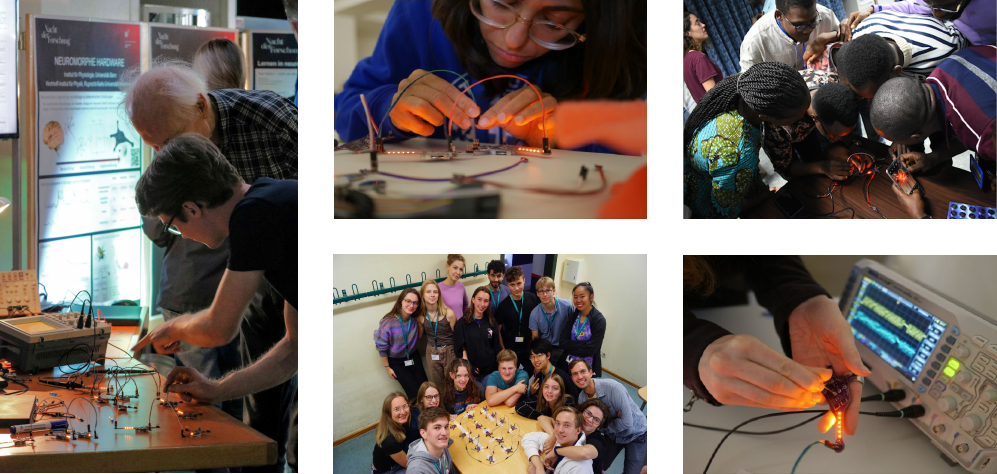}
    \caption{\label{fig:luiAction}%
		\boldmath\justifying%
		\Gls{lui} has played an integral role at various events all over the world for teaching and outreach applications:
		\foreignlanguage{ngerman}{Nacht der Forschung} (Switzerland,~2022), CapoCaccia Workshop toward Neuromorphic Intelligence (Italy,~2023), TReND in Africa (Ghana,~2023), and \foreignlanguage{ngerman}{Deutsche Schülerakademie} (Germany,~2023).
	}
\end{figure*}

\section{Exploring Neural Computation with Lu.i}\label{sec:didactic-application}

\Gls{lui} was designed to illustrate two of the fundamental aspects of biological neurons: spatio-temporal accumulation of input and event-based communication, both of which are captured by the \gls{lif} model.
These aspects can be demonstrated in a set of experiments of increasing complexity, some of them shown in~\cref{fig:all_the_luis}.

The first property -- leaky integration of input -- can be seen in \cref{fig:all_the_luis}A:
The membrane potential rises after weak excitatory stimuli and decays back to the resting potential, similarly with inhibitory input.
The resulting trajectories are shaped by the adjustable time constants $\tau_\text{syn}$ and $\tau_\text{mem}$.
These determine the time scales on which consecutive inputs are integrated and stacked.
Only when the threshold is reached, an efferent spike is triggered and visible externally.
On \gls{lui}, these dynamics can be observed using an on-board LED strip visualizing the membrane state and spike output, as shown in~\cref{fig:all_the_luis}A.
Neurons compute through this combination of analog integration and thresholding, for example by performing spatio-temporal coincidence detection.
Exploring the impact of the model parameters on this computation -- in case of coincidence detection on the sensitivity and detection window -- is a worthwhile educational exercise.

In contrast to the local computation on their membranes, neurons communicate through temporally sparse spike events.
This signal propagation can be demonstrated in a simple two-neuron network (\cref{fig:all_the_luis}C), where a synaptic connection is formed by a cable between the presynaptic axon and a postsynaptic dendrite.
By choosing a resting potential above the threshold, the first neuron can act as a regularly firing spike source to the second.
As before, the stacking of excitatory stimuli and the reset upon threshold crossing can be observed on the membrane of the postsynaptic cell.
The behavior of both neurons is clearly visible using the built-in LEDs without an external oscilloscope.
Already in this simple setup, the influence of the synaptic parameters can be explored:
For example, the combination of a short synaptic time constant and a strong excitatory weight can be used to trigger one spike for each incoming event.
Increasing the synaptic time constant, while lowering the weight, can lead to a delayed propagation of single spikes.
This can be used to build delay chains, which vividly illustrate the finite propagation speed of neural signals.
Once these chains are closed (\cref{fig:all_the_luis}D), their activity becomes self-sustained.

\Cref{fig:all_the_luis}E shows a more complex example, where rate-based \texttt{AND}, \texttt{OR} and -- in combination -- \texttt{XOR} gates are implemented using three \gls{lui} neurons.
In this case, the \texttt{OR} (\texttt{AND}) gate is implemented by a single neuron that has been tuned to fire for at least one (two) active presynaptic neurons.
The output of the \texttt{OR} neuron excites the \texttt{XOR} cell, with the \texttt{AND} neuron acting inhibitorily.

While the inputs \texttt{A} and \texttt{B} can be presented using \gls{lui} neurons (e.g., in leak-over-threshold configuration), we have used an external microcontroller to stimulate the network in \cref{fig:all_the_luis}E.
With a signal level of approximately \SI{2.5}{\volt}, \gls{lui}'s event output signal can be detected by most \SI{3.3}{\volt} and \SI{5}{\volt} microcontrollers.
The event inputs on \gls{lui} are compatible with signal levels from \SIrange{1.8}{20}{\volt}, allowing to interface with a great variety of sensors and devices.

Due to its simplicity, the \texttt{XOR} network is attractive in educational and outreach environments.
Inspired by existing literature, more complex networks have emerged from collaborations of researchers across all areas of neuroscience, including
realtime sound localization~\citep{jeffress1948place},
a balanced random network~\citep{brunel2000dynamics},
a ring attractor model~\citep{pisokas2020head},
an echo localization latch~\citep{wen2022curved}, and
-- with preprocessing of the analog signals -- a brightness change detection circuit.
\Gls{lui} has been used repeatedly to teach a younger audience about fundamentals of neuroscience and physical computing, especially in combination with a subsequent transition to neuromorphic research systems made accessible through EBRAINS\@.
Across all described applications, it was used to compellingly illustrate fundamental topics across a wide range of research areas from robotics to systems neuroscience.

\section{Discussion}\label{sec:discussion}

This manuscript presents \gls{lui}, a palm-sized electronic neuron with versatile applications for teaching and scientific outreach.
It can be used to illustrate the dynamics of individual neurons under different parametrizations and their interaction in small \acrlongpl{snn} (\cref{fig:all_the_luis}).
Featuring various connectivity options as well as on-board visualization aids, \gls{lui} can be used stand-alone or in combination with external equipment, like oscilloscopes, current sources, or microcontrollers.

\gls{lui} complements a range of pedagogical tools spanning from experimental to computational neuroscience~\citep{marzullo2012spiker,latimer2018open}.
Among those are guided experiments on tissue and living animals, which are arguably the most natural way to convey biological concepts but always imply ethical and logistical challenges.
Simulation-based curricula, on the contrary, trade immediacy with ease-of-use and simplicity, especially when considering graphical user interfaces~\citep{bekolay2014nengo,spreizer2021nestdesktop}.
To combine the advantages of both approaches, the concept of tangible hardware has been put forward before~\citep{eng2008first,pal2017single,baden2018spikeling,burdo2018neurobytes,renault2020neurino}.
As another effort in this direction, \gls{lui} combines an inviting interface with an analog yet accurate implementation of the \gls{lif} model.
The latter is sufficiently complex and flexible to allow illustration of fundamental biological phenomena as well as the concept of physical computation.
The \gls{pcb} is optimized for cost-effective manufacturing to ease acquisition especially for educational institutions.
With its engaging form factor, \gls{lui} has been welcomed at various conferences and workshops, leading to adoption by teachers and tutors in classrooms (\cref{fig:luiAction}).
As such, the project received enthusiastic responses initiating collaborations across both different areas of expertise and from pupils to faculty.

The \gls{lui} project is available as open hardware\footnote{\url{https://github.com/giant-axon/lu.i-neuron-pcb}} and undergoes active development.
The circuits are continuously improved and future versions might be accompanied by additional extensions, such as sensory spike sources or actuators.
In conjunction with the above-mentioned collaborations on courses and workshops using \gls{lui}, a curriculum of teaching material is being collected to nurture adoption among teaching personnel.

\section*{Acknowledgements}

The authors would like to thank
the Manfred Stärk Foundation for sponsoring the first batch of \gls{lui} neurons in preparation for the \foreignlanguage{ngerman}{Nacht der Forschung (Bern,  2022)},
Andreas Baumbach for his dedicated support during early development and prototyping,
Laura Kriener for patiently tutoring \gls{lui} users and passionately building functional networks,
Matthias Penselin and Nazarii Bogachuk for bringing the project into school classrooms,
the Human Brain Project for sponsoring subsequent production batches,
Björn Kindler for his administrative and public relations support,
Barbara Webb, Ben Efron, Chenxi Wen, Robin Dietrich, Sarah Hamburg, Shreyan Banerjee, and Yulia Sandamirskaya for pioneering experiments during the CapoCaccia Workshop toward Neuromorphic Intelligence 2023, %
Saray Soldado-Magraner for taking \gls{lui} to Ghana and her photos thereof,
and
the \foreignlanguage{ngerman}{Deutsche Schülerakademie} for kindly authorizing the use of a group photo.
Finally, we'd like to thank the NeuroTMA and ElectronicVision(s) groups for their continuous support and inspirational atmosphere.

\section*{Funding}

The presented work has received funding from
the Manfred Stärk Foundation,
the EC Horizon 2020 Framework Programme under grant agreement 945539 (HBP) and Horizon Europe grant agreement 101147319 (EBRAINS 2.0),
the \foreignlanguage{ngerman}{Deutsche Forschungsgemeinschaft} (DFG, German Research Foundation) under Germany's Excellence Strategy EXC 2181/1-390900948 (Heidelberg STRUCTURES Excellence Cluster).

\printbibliography

\end{document}